%% file: main.tex
\documentclass[sigplan,screen,,timestamp]{acmart}
\AtBeginDocument{%
  }

\usepackage{stmaryrd}
\usepackage[scr=boondoxo]{mathalpha}
\usepackage{relsize}
\usepackage{mathpartir}

\usepackage{listings}
\lstdefinelanguage{Koka}{
  morekeywords={type,effect,ctl,fun,match,handler,with,resume,div,exn,pure,if,then,else,val},
  sensitive=true,
  morecomment=[l]{//},
  morecomment=[s]{/*}{*/},
  morestring=[b]",
  morestring=[b]',
  alsoletter={:}
}
\newcommand{\lstbasicstyle}{\ttfamily\ttfamily\small}
\newcommand{\lstkeywordstyle}{\bfseries}
\newcommand{\lstcommentstyle}{\sl}
\newcommand{\lstnumberstyle}{\scriptsize\em}
\lstdefinestyle{default}{%
  backgroundcolor=\color{white},%
  basicstyle=\lstbasicstyle,%
  commentstyle=\lstcommentstyle,%
  keywordstyle=\lstkeywordstyle,%
  columns=fullflexible,%
  keepspaces=true,%
  mathescape=true%
}
\lstdefinestyle{number}{%
  numbers=left,%
  numberstyle=\lstnumberstyle,%
  xleftmargin=2em%
}
\usepackage{xcolor}

\definecolor{Cayenne}{HTML}{941100}
\definecolor{Mocha}{HTML}{945200}
\definecolor{Asparagus}{HTML}{929000}
\definecolor{Fern}{HTML}{4F8F00}
\definecolor{Clover}{HTML}{008F00}
\definecolor{Moss}{HTML}{009051}
\definecolor{Teal}{HTML}{009193}
\definecolor{Ocean}{HTML}{005493}
\definecolor{Midnight}{HTML}{011993}
\definecolor{Eggplant}{HTML}{011993}
\definecolor{Plum}{HTML}{942193}
\definecolor{Maroon}{HTML}{942193}

\setcopyright{none}
\copyrightyear{2025}
\acmYear{2025}
\acmDOI{XXXXXXX.XXXXXXX}
\acmConference[]{}{}{}

\input{macros}

\begin{document}

\title{Towards Cumulative Abstract Semantics via Handlers}
\subtitle{Short Paper}

\author{Cade Lueker}
\affiliation{
  \institution{University of Colorado Boulder}
  \country{USA}
}
\email{cade.lueker@colorado.edu}

\author{Andrew Fox}
\affiliation{
  \institution{University of Colorado Boulder}
  \country{USA}
}
\email{andrew.fox@colorado.edu}

\author{Bor-Yuh Evan Chang}
\affiliation{
  \institution{University of Colorado Boulder}
  \country{USA}
}
\additionalaffiliation{
  \institution{Amazon}
  \country{USA}.
  Bor-Yuh Evan Chang holds concurrent appointments at the University of Colorado Boulder and as an Amazon Scholar. This paper describes work performed at the University of Colorado Boulder and is not associated with Amazon.
}
\email{evan.chang@colorado.edu}

\begin{abstract}
\input{sections/Abstract}
\end{abstract}

\maketitle

\section{Introduction}
\input{sections/Intro}

\section{Cumulative Abstract Semantics}\label{sec:overview}
\input{sections/RecReduceOverview}

\section{Related Work}
\input{sections/RelatedWorks}

\section{Conclusion}
\input{sections/Conclusion}

\bibliographystyle{ACM-Reference-Format}
\bibliography{ref}

\end{document}

%% file: macros.tex
\newenvironment{displaysmall}
  {\begin{center}\begin{minipage}{0.96\linewidth}\centering\small}
  {\end{minipage}\end{center}}
\newenvironment{grammar}
  {\begin{array}{rr@{}c@{}lrl}}
  {\end{array}}
\newenvironment{mathparsmall}
  {\begin{displaysmall}\begin{mathpar}}
  {\end{mathpar}\end{displaysmall}}
\newcommand{\bnfdef}{\mathrel{::=}}
\newcommand{\bnfalt}{\mid}

\newcommand{\hasty}{\colon}
\newcommand{\mto}{\shortrightarrow}
\newcommand{\marrow}{\rightharpoonup}
\newcommand{\colorcode}[1]{{\color{Cayenne}#1}}
\newcommand{\colorelim}[1]{{\color{Teal}#1}}
\newcommand{\colorintro}[1]{{\color{Maroon}#1}}
\newcommand{\colorlower}[1]{{\color{Eggplant}#1}}
\newcommand{\colorunsubt}[1]{{\color{Fern}#1}}
\newcommand{\colorsubt}[1]{{\color{Asparagus}#1}}
\newcommand{\fmtmvar}[1]{\mathit{#1}}
\newcommand{\fmtmkw}[1]{\mathrm{#1}}
\newcommand{\fmtmtyp}[1]{\mathrm{#1}}
\newcommand{\fmtmtyvar}[1]{\mathscr{#1}}
\newcommand{\fmtcode}[1]{\colorcode{\text{\tt #1}}}
\newcommand{\fmtcodekw}[1]{\colorcode{\text{\relsize{-1}\tt #1}}}

\newcommand{\fmtelim}[1]{\colorelim{\acute{#1}}}
\newcommand{\fmtelimkw}[1]{\fmtelim{\fmtmkw{#1}}}
\newcommand{\fmtintro}[1]{\colorintro{\grave{#1}}}
\newcommand{\fmtintrokw}[1]{\fmtintro{\fmtmkw{#1}}}
\newcommand{\fmtlower}[1]{\colorlower{\grave{#1}}}
\newcommand{\fmtlowerkw}[1]{\fmtlower{\fmtmkw{#1}}}
\newcommand{\fmtunsubt}[1]{\colorunsubt{\breve{#1}}}
\newcommand{\fmtsubt}[1]{\colorsubt{#1}}
\newcommand{\expr}{e}
\newcommand{\Expr}{\fmtmtyp{expr}}
\newcommand{\ExprE}{\fmtelim{\Expr}}
\newcommand{\num}{n}
\newcommand{\Num}{\fmtmtyp{int}}

\newcommand{\edom}{d}
\newcommand{\edomtyvar}{\fmtmtyvar{\edom}}

\newcommand{\effwild}{\mathunderscore}
\newcommand{\effsep}{,}

\newcommand{\polyeff}[2]{#2 \mid #1}
\newcommand{\effty}[1]{\langle #1 \rangle}
\newcommand{\fnty}[2]{#1 \marrow #2}

\newcommand{\effretty}[2]{\effty{#1}\;#2}

\newcommand{\polyeffretty}[2]{\effretty{\polyeff{\effwild}{#1}}{#2}}
\newcommand{\ctlhkw}{\fmtmkw{ctl}}
\newcommand{\ctlhty}[2]{\ctlhkw\;#1 \mto #2}
\newcommand{\funhkw}{\fmtmkw{fun}}
\newcommand{\funhty}[2]{\funhkw\;#1 \mto #2}
\newcommand{\intronum}{\fmtintro{\num}}
\newcommand{\numinf}{\fmtmvar{bnd}}
\newcommand{\minterval}{r}
\newcommand{\mInterval}{\fmtmtyp{interval}}
\newcommand{\mkinterval}[2]{[#1, #2]}
\newcommand{\ninf}{-\infty}
\newcommand{\pinf}{+\infty}
\newcommand{\binexpr}[3]{#1 \mathbin{#2} #3}
\newcommand{\binctlhty}[3]{\ctlhty{#1 \mto #2}{#3}}
\newcommand{\binfunhty}[3]{\funhty{#1 \mto #2}{#3}}

\newcommand{\plusop}{+}
\newcommand{\jsyplusop}{\fmtcode{\plusop}}
\newcommand{\jsyplus}[2]{\binexpr{#1}{\jsyplusop}{#2}}

\newcommand{\elimplusop}{\fmtelim{\plusop}}
\newcommand{\elimplus}[2]{\binexpr{#1}{\elimplusop}{#2}}
\newcommand{\introplusop}{\fmtintro{\plusop}}

\newcommand{\evalop}{\Downarrow}
\newcommand{\unsubtevalop}{\fmtunsubt{\evalop}}
\newcommand{\subtevalop}{\fmtsubt{\evalop}}
\newcommand{\jsubteval}[2]{#1 \mathrel{\subtevalop} #2}
\newcommand{\ifnzkw}{ifnz}
\newcommand{\thenkw}{then}
\newcommand{\elsekw}{else}
\newcommand{\jsyifnzkw}{\fmtcodekw{\ifnzkw}}
\newcommand{\jsythenkw}{\fmtcodekw{\thenkw}}
\newcommand{\jsyelsekw}{\fmtcodekw{\elsekw}}
\newcommand{\jsyifnz}[3]{\jsyifnzkw \, #1 \, \jsythenkw \, #2 \, \jsyelsekw \, #3}
\newcommand{\elimifnzkw}{\fmtelimkw{\ifnzkw}}
\newcommand{\elimthenkw}{\colorelim{\fmtmkw{\thenkw}}}
\newcommand{\elimelsekw}{\colorelim{\fmtmkw{\elsekw}}}
\newcommand{\elimifnz}[3]{\elimifnzkw \, #1 \, \elimthenkw \, #2 \, \elimelsekw \, #3}
\newcommand{\introifnzkw}{\fmtintrokw{\ifnzkw}}

\newcommand{\terctlhty}[4]{\ctlhty{#1 \mto #2 \mto #3}{#4}}
\newcommand{\terfunhty}[4]{\funhty{#1 \mto #2 \mto #3}{#4}}

\newcommand{\assumenzkw}{assume\mathunderscore{}nz}
\newcommand{\assumezkw}{assume\mathunderscore{}z}
\newcommand{\introassumenzkw}{\fmtlowerkw{\assumenzkw}}
\newcommand{\introassumezkw}{\fmtlowerkw{\assumezkw}}
\newcommand{\introassumenz}[2]{\introassumenzkw\, #1 \, #2}
\newcommand{\introassumez}[2]{\introassumezkw\, #1 \, #2}
\newcommand{\joinop}{\sqcup}
\newcommand{\introjoinop}{\fmtlower{\joinop}}



%% file: sections/Abstract.tex
We consider the problem of modularizing control flow in a generic abstract interpretation framework.
A generic abstract interpretation framework is not truly flexible if it does not allow interpreting with different path- and flow-sensitivities, by going forwards or backwards, and over- or under-approximately.
Most interpreters inherently intertwine syntax and semantics, making the implementation antagonistic to modularity. 
Current approaches to modular designs require the use of complex data structures (e.g., monad transformers), providing modularity but often proving unwieldy (e.g., lifts).
We observe that leveraging algebraic effects within an interpreter facilitates the accumulation of semantic fragments against a fixed syntax.
In this paper, we define \emph{cumulative abstract semantics}, illustrating the potential for creating multiple dynamic evaluators and static analyses from one interpreter.
This modularity is achieved by grouping effects into two categories: syntax \emph{elimination} and domain-semantic \emph{introduction} handlers. 
Our contribution shows the benefits of using effects as an instrument for designing a clean, elegant, and modular abstract interpretation framework.

%% file: sections/Intro.tex
Abstract interpretation frameworks are often burdened with the delicate balance of modularity while being simple to extend.
In part, this challenge is an instance of the well-known expression problem, to which there are many solutions.
Numerous approaches offer modularity at the price of simplicity.
This dilemma often means developers will opt for the less modular approach, leading to the typical intertwined fusion of syntax and semantics. 

Dynamic evaluators and static analyzers are valuable, demanding both time and effort to implement. 
If an interpreter is too delicate and monolithic, despite its value, it will ultimately be abandoned.
This leaves most of its functionality to be re-implemented.
The ideal abstract interpretation framework would not only be modular but require little overhead to extend. 
Being forced to abandon an implementation due to new requirements of the semantic domain and finding it simpler to start from the ground up prevents progress and leads to lost time.
This has led to many abstract analysis frameworks losing momentum once completed, as they cannot easily extend new features or domains.

Interpreters serve the function of giving semantics to syntax, meaning separating the two is a difficult task.
One existing approach is through monads, but the complexity often prevents the full realization of monadic modularity~\cite{Brady13}.
Existing modular monadic approaches use tightly coupled components, traversed with monad transformers, making it a complex task to modify elements in isolation~\cite{Schrijvers19, Liang95}. Monadic encapsulation is represented by layers of stacked monads, each handling specific functionality, relying on monad transformers to couple distinct components.
This structure means computations must be manually threaded through adjacent layers.

Recently, a new approach to functional encapsulation has emerged in this space: effects and effect handlers~\cite{Plotkin_2013, Brady13, Poulsen23}.
Several new languages natively support these effects systems, such as Koka~\cite{leijen14, leijen17} and Flix~\cite{Madsen16}.
While these languages have explored effect systems for language design and extension, their potential for modularizing abstract interpretation frameworks has remained largely uncharted.
At a high level, the distinction between an effect-oriented interface and a monad stack is the ease of composition.
However, languages with first-class, polymorphic effects, such as Koka, can infer a triggered effect's return type and call the corresponding handler automatically~\cite{leijen14}.
Consequently, effect systems allow for the separation of syntax and semantics with effect signatures and effect handlers; to do this, they rely on delimited continuations within handlers, implemented to fill out the type requirements of the effects they handle~\cite{Hillerstrom16}. 
It is important to note that dynamically scoped handlers can cause continuations to capture more than they should when an effect resumes. Grouping effectful operations with multiple resumptions into one effect interface ensures each operations' handler is in the same scope. As a result, each continuation captures precisely what it should at every resumption.
By varying the order of applied resumptions in handlers for the same effects, control flow can be transformed~\cite{Hillerstrom16, leijen17}.
As a result the same effectful skeleton or effect interface can express single-path or multiple-path interpretations by changing only one handler.

In this paper, we show promise for defining \textit{cumulative abstract semantics}, developing a lexicon of abstract-semantic handlers substantiating executable abstract interpreters. In particular, our contributions are as follows:
\begin{itemize}
  \item We observe that parameterizing an interpreter by an abstract domain using effect types corresponds to defining an effect interface consisting of \emph{introduction handlers} (\autoref{sec:introduction-handlers}). Introduction handlers enable a cumulative definition of abstract domain operations.
  \item We go beyond fixed control flow for interpretation by proposing \emph{elimination handlers} (\autoref{sec:elimination-handlers}). Elimination handlers eliminate syntax and leverage multiple resumptions to cumulatively define domain-specific semantics with different control flow.
\end{itemize}

%% file: sections/RecReduceOverview.tex
To convey the value of using handlers as the foundation for an abstract interpretation framework, we illustrate our approach with a small, end-to-end example on a minimal expression language with integer addition and conditionals. Let us first consider addition (without conditionals):
\begin{mathparsmall}\begin{grammar}
\text{\emph{e}xpressions} & \expr & \hasty & \Expr & \bnfdef &
  \num
  \bnfalt \jsyplus{\expr_1}{\expr_2}
\\
\text{i\emph{n}tegers} & \num & \hasty & \Num
\end{grammar}\end{mathparsmall}
In particular, we consider a progressive refactoring of a non-compositional concrete interpreter for expressions $\expr$ into a \emph{cumulative abstract interpreter}.

Let us write $\jsubteval{\expr}{\num}$ for the judgment form that says, "Expression $\expr$ evaluates to integer $\num$," which is the form of a standard, big-step operational semantics.  We call an implementation of this judgment form, a \emph{substantiated} interpreter, as it is a executable function --- all effects that must be handled have corresponding handlers, allowing the interpreter to execute programs:
\begin{mathparsmall}\begin{grammar}
\text{substantiated evaluation} & \subtevalop & \hasty & \fnty{\Expr}{\Num}
\end{grammar}\end{mathparsmall}
That is, it may have core language effects, inferred by Koka like divergence \lstinline!div! and exceptions \lstinline!exn!, however these effects do not require handling allowing their presence to preserve our notion of substantiation.

The above corresponds to something like the following code snippet in Koka wherein \lstinline!pure! denotes the lack of \lstinline!div! and \lstinline!exn! effects:
\begin{displaysmall}\begin{lstlisting}
type expr
  Int(n: int)
  Plus(e1: expr, e2: expr)
fun eval(e: expr): pure int
\end{lstlisting}\end{displaysmall}
For presentation, we incorporate our on-paper notation into our code snippets where the context is clear; for example, consider the above code snippet with our on-paper notation:
\begin{displaysmall}\begin{lstlisting}
type $\Expr$
  $\num$
  $\jsyplus{\expr_1}{\expr_2}$
fun $\subtevalop$($\expr$: $\Expr$): pure $\Num$
\end{lstlisting}\end{displaysmall}

\begin{figure}[t]\small\begin{lstlisting}
fun $\subtevalop_0$($\expr$: $\Expr$): $\Num$
  match $\expr$
    $\num$ -> $\num$
    $\jsyplus{\expr_1}{\expr_2}$ -> $\subtevalop_0$($\expr_1$) + $\subtevalop_0$($\expr_2$)
\end{lstlisting}
\caption{A standard, big-step, concrete substantiated interpreter $\subtevalop_0$.}
\label{fig:eval-v0}
\end{figure}
To start, consider a monolithic concrete interpreter $\subtevalop_0$ for the language in \autoref{fig:eval-v0}.
This interpreter is not \emph{cumulative} because it does not expose any of its internal evaluation structure for reuse or extension to define new --- concrete or abstract --- semantics.

\subsection{Domain Parametrization and Introduction Handlers}
\label{sec:introduction-handlers}

Following abstract interpretation~\cite{Cousot77}, we first refactor the concrete interpreter $\subtevalop_0$ from \autoref{fig:eval-v0} into an abstract interpreter by \emph{parametrizing} it over an \emph{evaluation domain}:
\begin{mathparsmall}\begin{grammar}
\text{evaluation \emph{d}omains} & \edom & \hasty & \edomtyvar
\end{grammar}\end{mathparsmall}
We write $\edomtyvar$ for the type variable representing the evaluation domain.
Wherever we write $\edomtyvar$ in an effect type or function type, we mean that it is parameterized (i.e., generic) on the given evaluation domain.

\begin{figure}[b]\small\begin{lstlisting}
effect<$\edomtyvar$> fun $\intronum$($\num$: $\Num$): $\edomtyvar$
effect<$\edomtyvar$> fun $\introplusop$($\edom_1$: $\edomtyvar$, $\edom_2$: $\edomtyvar$): $\edomtyvar$
$\rule{0pt}{3ex}$fun $\unsubtevalop_1$($\expr$: $\Expr$): $\polyeffretty{\intronum \effsep \introplusop}{\edomtyvar}$
  match $\expr$
    $\num$ -> $\intronum$
    $\jsyplus{\expr_1}{\expr_2}$ -> $\unsubtevalop_1$($\expr_1$) $\introplusop$ $\unsubtevalop_1$($\expr_2$)
$\rule{0pt}{3ex}$fun $\subtevalop_2$($\expr$: $\Expr$): pure $\Num$
  with fun $\intronum$($\num$: $\Num$) $\num$
  with fun $\introplusop$($\num_1$: $\Num$, $\num_2$: $\Num$) $\num_1$ + $\num_2$
  $\unsubtevalop_1$($\expr$)
$\rule{0pt}{3ex}$fun $\subtevalop_3$($\expr$: $\Expr$): pure $\mInterval$
  with fun $\intronum$($\num$: $\Num$) $\mkinterval{\num}{\num}$
  with fun $\introplusop$($\minterval_1$: $\mInterval$, $\minterval_2$: $\mInterval$) $\ldots$
  $\unsubtevalop_1$($\expr$)
\end{lstlisting}
\caption{An abstract unsubstantiated interpreter $\unsubtevalop_1$, a concrete substantiated interpreter $\subtevalop_2$, and an abstract substantiated interpreter with the interval domain $\subtevalop_3$.}
\label{fig:eval-v1}
\end{figure}

We define domain operations as handlers for an effect interface:
\begin{mathparsmall}\begin{grammar}
\text{integer literal introduction} & \intronum & \hasty & \funhty{\Num}{\edomtyvar}
\\
\text{plus introduction} & \introplusop & \hasty & \binfunhty{\edomtyvar}{\edomtyvar}{\edomtyvar}
\end{grammar}\end{mathparsmall}
where we write $\funhkw$ for a tail-resumptive handler (i.e., one that resumes with its final value using the same keyword as Koka).
The $\intronum$ handler represents integer literals in the domain, and the $\introplusop$ handler combines two domain elements via an abstract addition operation.
In \autoref{fig:eval-v1}, we show an unsubstantiated interpreter $\unsubtevalop_1$ that is generic over the evaluation domain $\edomtyvar$ by using the effect interface with $\intronum$ and $\introplusop$.
We can define a concrete substantiated interpreter $\subtevalop_2$ by providing handlers for the domain operations that implement integer addition \lstinline!+! in the meta language (i.e., Koka).
Note that the unsubstantiated interpreter $\unsubtevalop_1$ is simply the reduce, or fold, over the expression type $\Expr$. This reduce uses effect handlers that we can see as either the algebra or as semantic domain operations.

We can instantiate this interpreter with different domains. For example, consider the interval abstact domain:
\begin{mathparsmall}\begin{grammar}
\text{inte\emph{r}vals} & \minterval & \hasty & \mInterval & \bnfdef & \mkinterval{\numinf_1}{\numinf_2}
\\
\text{interval \emph{b}ou\emph{nd}s} & \numinf &&& \bnfdef & \ninf \bnfalt \num \bnfalt \pinf
\end{grammar}\end{mathparsmall}
that represents sets of integers as intervals $\minterval \bnfdef \mkinterval{\numinf_1}{\numinf_2}$ consisting of a lower bound $\numinf_1$ and an upper bound $\numinf_2$.
Implementing the $\intronum$ handler corresponds to implementing the representation function (i.e., $\beta$) that maps concrete integers to intervals, and implementing the $\introplusop$ handler corresponds to implementing the abstract addition operation over intervals.

We call such handlers that introduce domain elements \emph{introduction handlers} and correspond to introduction forms of the syntax.

\subsection{Interpretation Parametrization and Elimination Handlers}
\label{sec:elimination-handlers}

The unsubstantiated interpreter $\unsubtevalop_1$ from \autoref{fig:eval-v1} is still not cumulative because it does not expose its internal evaluation structure: it bakes in the control flow of evaluating the plus expression $\jsyplus{\expr_1}{\expr_2}$ (i.e., the order of evaluating sub-expressions $\expr_1$ and $\expr_2$ and then combining their results). Worse, this control flow is fixed for all domains.

\begin{figure}[b]\small\begin{lstlisting}
fun $\subtevalop_0$($\expr$: $\Expr$): $\Num$
  match $\expr$
    $\ldots$
    $\jsyifnz{\expr_1}{\expr_2}{\expr_3}$ ->
      if $\subtevalop_0$($\expr_1$)$\,$!=$\,$0 then $\subtevalop_0$($\expr_2$) else $\subtevalop_0$($\expr_3$)
\end{lstlisting}
\caption{Extending the standard, big-step, concrete substantiated interpreter $\subtevalop_0$ with conditionals.}
\label{fig:eval-v0-if}
\end{figure}

Consider extending the language with conditionals:
\begin{mathparsmall}\begin{grammar}
\text{\emph{e}xpressions} & \expr & \hasty & \Expr & \bnfdef & \cdots
  \bnfalt \jsyifnz{\expr_1}{\expr_2}{\expr_3}
\end{grammar}\end{mathparsmall}
where $\jsyifnz{\expr_1}{\expr_2}{\expr_3}$ evaluates $\expr_1$ and if the result is non-zero, evaluates $\expr_2$; otherwise, it evaluates $\expr_3$.
In a standard big-step concrete interpreter, the control flow of interpreting the syntax $\jsyifnz{\expr_1}{\expr_2}{\expr_3}$ will only evaluate one of the branches $\expr_2$ or $\expr_3$ as shown in \autoref{fig:eval-v0-if}.
Extending the unsubstantiated interpreter $\unsubtevalop_1$ to handle conditionals with an $\introifnzkw$ introduction handler
\begin{mathparsmall}\begin{grammar}
  \text{ifnz introduction} & \introifnzkw & \hasty & \terfunhty{\edomtyvar}{\edomtyvar}{\edomtyvar}{\edomtyvar}
\end{grammar}\end{mathparsmall}
would not seem to support this short-circuiting control flow.
At the same time, an abstract interpreter would typically need to evaluate both branches and then join the results to soundly approximate the conditional expression.
A cumulative interpreter should be able to support both control flows.

To make our interpreter cumulative, we refactor $\unsubtevalop_1$ to expose its internal evaluation structure using \emph{elimination handlers}:
\begin{mathparsmall}\begin{grammar}
\text{plus elimination} & \elimplusop & \hasty & \binctlhty{\Expr}{\Expr}{\Expr}
\\
\text{ifnz elimination} & \elimifnzkw & \hasty & \terctlhty{\Expr}{\Expr}{\Expr}{\Expr}
\end{grammar}\end{mathparsmall}
that are general control effects and correspond to pattern matching or elimination forms of the syntax.

\begin{figure}\small\begin{lstlisting}
$\rule{0pt}{3ex}$effect $\ExprE$
  ctl $\elimplusop$($\expr_1$: $\Expr$, $\expr_2$: $\Expr$): $\Expr$
  ctl $\elimifnzkw$($\expr_1$: $\Expr$, $\expr_2$: $\Expr$, $\expr_3$: $\Expr$): $\Expr$
$\rule{0pt}{3ex}$fun $\unsubtevalop_4$($\expr$: $\Expr$): $\polyeffretty{\intronum \effsep \elimplusop \effsep \elimifnzkw}{\edomtyvar}$
  match $\expr$
    $\num$ -> $\intronum$
    $\jsyplus{\expr_1}{\expr_2}$ -> $\unsubtevalop_4$($\elimplus{\expr_1}{\expr_2}$)
    $\jsyifnz{\expr_1}{\expr_2}{\expr_3}$ -> $\unsubtevalop_4$($\elimifnz{\expr_1}{\expr_2}{\expr_3}$)
$\rule{0pt}{3ex}$fun $\subtevalop_5$($\expr$: $\Expr$): pure $\Num$
  with fun $\intronum$($\num$: $\Num$) $\num$
  with fun $\introplusop$($\num_1$: $\Num$, $\num_2$: $\Num$) $\num_1$ + $\num_2$
  with handler
    ctl $\elimplusop$($\expr_1$: $\Expr$, $\expr_2$: $\Expr$)
      resume($\expr_1$) $\introplusop$ resume($\expr_2$)
    ctl $\elimifnzkw$($\expr_{guard}$: $\Expr$, $\expr_{true}$: $\Expr$, $\expr_{false}$: $\Expr$)
      if resume($\expr_1$)$\,$!=$\,$0 then resume($\expr_2$) else resume($\expr_3$)
  $\unsubtevalop_4$($\expr$)
\end{lstlisting}
\caption{An abstract interpreter $\unsubtevalop_4$ unsubstantiated by the introduction handler $\intronum$ and elimination handlers $\elimplusop, \elimifnzkw$ and a concrete substantiated interpreter $\subtevalop_5$.}
\label{fig:eval-v2}
\end{figure}

We can now refactor the unsubstantiated interpreter $\unsubtevalop_1$ from \autoref{fig:eval-v1} to expose its internal evaluation structure for reuse and extension, which we show as $\unsubtevalop_4$ in \autoref{fig:eval-v2}.
Observe that $\unsubtevalop_4$ simply calls the corresponding elimination handler for plus expressions and conditional expressions to select the next step of evaluation.
As a general control effect, elimination handlers can use multiple resumptions to evaluate multiple (sub-)expressions, as we show in the concrete substantiated interpreter $\subtevalop_5$.

\begin{figure}\small\begin{lstlisting}
effect<$\edomtyvar$> fun $\introifnzkw$($\edom_1$: $\edomtyvar$, $\edom_2$: $\edomtyvar$, $\edom_3$: $\edomtyvar$): $\edomtyvar$
$\rule{0pt}{3ex}$fun $\unsubtevalop_6$($\expr$: $\Expr$): $\polyeffretty{ \intronum \effsep \introplusop }{\edomtyvar}$
  with fun $\introassumenzkw$($\edom_1$: $\edomtyvar$, $\edom_2$: $\edomtyvar$) $\ldots$
  with fun $\introassumezkw$($\edom_1$: $\edomtyvar$, $\edom_3$: $\edomtyvar$) $\ldots$
  with fun $\introjoinop$($\edom_2$: $\edomtyvar$, $\edom_3$: $\edomtyvar$) $\ldots$
  with fun $\introifnzkw$($\edom_1$: $\edomtyvar$, $\edom_2$: $\edomtyvar$, $\edom_3$: $\edomtyvar$)
    ($\introassumenz{\edom_1}{\edom_2}$) $\introjoinop$ ($\introassumez{\edom_1}{\edom_3}$)
  with handler
    ctl $\elimplusop$($\expr_1$: $\Expr$, $\expr_2$: $\Expr$)
      resume($\expr_1$) $\introplusop$ resume($\expr_2$)
    ctl $\elimifnzkw$($\expr_{guard}$: $\Expr$, $\expr_{true}$: $\Expr$, $\expr_{false}$: $\Expr$)
      $\introifnzkw$(resume($\expr_{guard}$), resume($\expr_{true}$), resume($\expr_{false}$))
  $\unsubtevalop_4$($\expr$)
\end{lstlisting}
\caption{An abstract interpreter $\unsubtevalop_6$ defined in terms of $\unsubtevalop_4$ from \autoref{fig:eval-v2}, which is now unsubstantiated only by the introduction handlers $\intronum, \introplusop$}
\label{fig:eval-v3}
\end{figure}

Finally, we can define an abstract interpreter $\unsubtevalop_6$ in \autoref{fig:eval-v3} that, unlike the concrete interpreter $\subtevalop_5$, interprets conditionals by evaluating both branches and then joining their results.
To emphasize this, we use an introduction handler for conditionals $\introifnzkw$ that is defined in terms of \emph{lowering handlers} $\introassumenzkw$ for assuming the guard being non-zero, $\introassumezkw$ for assuming the guard being zero, and $\introjoinop$ for joining to abstract domain elements.
Observe that we can see $\introassumenzkw, \introassumezkw,$ and $\introjoinop$ as standard abstract domain operations to define the semantics of conditionals. Additionally, because these effects only introduce new transformations to values in our domain, without interacting with syntax by reducing expressions, they do not need to be handled for every analysis. Instead, they can be used in a demanded fashion, only when needed, emphasizing the \textit{cumulative} nature of our design.

We can see the unsubstantiated interpreter $\unsubtevalop_6$ as a \emph{cumulative abstract interpreter} from $\unsubtevalop_4$ because it is defined in terms of $\unsubtevalop_4$. We can further see that $\unsubtevalop_6$ can be substantiated by the interval abstract domain or any other abstract domain supporting the required domain operations $\intronum, \introplusop$.

This framework also allows changes in control with reuse in the rest of the handlers. For example, a backwards interpretation would only require changing the plus elimination handler to the following:
\begin{lstlisting}
 with handler
    ctl $\elimplusop$($\expr_1$: $\Expr$, $\expr_2$: $\Expr$)
      resume($\expr_2$) $\introplusop$ resume($\expr_1$)
\end{lstlisting}

This new handler evaluates the right hand side before the left hand side, changing the order of evaluation and allowing reuse of the other three handlers.

%% file: sections/RelatedWorks.tex
\paragraph{Effect-Based Interpretation}
To represent effects in Has\-kell, Kiselyov championed the use of free monads, while also allowing the composition of computations by taking their co-product~\cite{Kiselyov15}.
As a result, evaluation was represented as a fold over the co-product type signature of every effect. This facilitated an extensible effect system, at the cost of manually lifting and injecting each matched effect type. The idea to make an effect based interpreter was theorized and toyed with by \citet{Reinders23}, but never fully implemented until very recently. In \citet{BunkenBurg24} started using effect handlers embedded in Haskell to build a modular interpreter, one of the first of its kind. However, its focus was on concrete interpretation rather than abstract interpretation. This work highlights the potential of embedded effects for modular interpreters, which our approach extends to abstract domains and abstract semantics. The Semantic project  combined various functional programming components to create a multilingual code differential analysis of GitHub repositories, relying on algebraic effects \cite{Thomson22, github_semantic}.

\paragraph{Modular Abstract Interpreters}
\citet{Keidel18} used arrow transformers from category theory to compose abstract domains while preserving soundness 
in the Sturdy abstract interpretation framework.
Their approach provides soundness guarantees for free but is fixed to non-relational domains and requires an implementation of a monolithic, generic interpreter for every language.
\citet{Sergey13} demonstrated the ability of monads to modularize semantics, allowing them to be combined in unique ways for various interpretations similar to our technique. However this monadic approach carries the same burden of threading computations and context through a dense monadic stack of operations.
\citet{Michelland24} advanced the usage of monadic modularity with their ITree-based framework for abstract interpretation, composing state and control flow monads for sound meta-theory in Roq. Their approach supports modular analyses (e.g., binding-time analysis) but requires complex transformer stacks. We seek to expand on this idea with our effect based skeleton, which relies on Koka's handlers for seamless composition single and multi-path analyses, aiming for easier integration.

%% file: sections/Conclusion.tex
We presented the groundwork for a modular abstract interpretation framework that leverages algebraic effects for cumulative semantic definition.
Unlike monadic interpreters requiring manual plumbing, row-poly\-morph\-ic effects automate handler selection, allowing for a seamless sequencing of effectful domain-specific interpretations.

By thinking of effects as either eliminating syntax or introducing domain values, we were able to define an unsubstantiated evaluation function that when paired with different handlers allows for not only varied semantic domains but also varied interpretation control flow.
This approach allows for a high level of reuse as the syntax is fixed, while the domain-specific implementation can be swapped, modified, or expanded upon with minimal boilerplate.
Our \emph{cumulative abstract semantics} allow us to recycle handlers for reuse as new domains are added.
Our framework enables single-path and multi-path control flow with minimal code changes.
This methodology allows for the cumulation of domains and semantics supported by our framework, as every addition of handlers provides new functionality while building on previous designs.
We plan to continue building on this technique into a fully-fledged abstract interpretation framework --- supporting the syntax of a richer language and providing several substantiated interpreters with various concrete and abstract domains.